\documentclass[a4paper,fleqn]{cas-dc}

\usepackage[authoryear,longnamesfirst]{natbib}
\usepackage[utf8]{inputenc}
\usepackage{mhchem}
\def\tsc#1{\csdef{#1}{\textsc{\lowercase{#1}}\xspace}}
\tsc{WGM}
\tsc{QE}
\tsc{EP}
\tsc{PMS}
\tsc{BEC}
\tsc{DE}

\begin{document}
\let\WriteBookmarks\relax
\def\floatpagepagefraction{1}
\def\textpagefraction{.001}

\shorttitle{Li|Li$_{3}$OCl interface mechanisms for solid state Li-battery}


\title [mode = title]{Electrochemical stability and lithium insertion at the Li|Li$_{3}$OCl solid electrolyte interface}                      


%
\author[1]{Deobrat Singh}

\cormark[1]


\ead{deosing@kth.se}


\credit{Conceptualization of this study, Methodology, Software}

\affiliation[1]{organization={Department of Materials Science and Engineering, Royal Institute of Technology}, 
    city={Stockholm},
    postcode={SE-100 44}, 
    country={Sweden}}

\author[1]{Li-Yun Tian}

\author[2,3]{Moyses Araujo}


\affiliation[2]{organization={Department of Engineering and Physics, Karlstad University},
    city={Karlstad},
    postcode={651 88}, 
    country={Sweden}}

\affiliation[3]{organization={Department of Physics and Astronomy, Division of Materials Theory, Uppsala University},
    city={Uppsala},
    postcode={Box 516, SE-75120}, 
    country={Sweden}}

\author%
[1,3]
{Raquel Liz\'arraga}
\cormark[2]
\ead{raqli@kth.se}

\affiliation[4]{organization={Wallenberg Initiative Materials Science for Sustainability, Department of Materials Science and Engineering, The Royal Institute of Technology},
    city={Stockholm},
    postcode={SE-10044},
    country={Sweden}}

\cortext[cor1]{Corresponding author}
\cortext[cor2]{Principal corresponding author}



\begin{abstract}
Solid-state lithium batteries have attracted considerable attention due to their potential to provide improved safety and higher energy density compared with conventional liquid electrolyte batteries. However, the stability of the interface between Li metal anodes and solid electrolytes remains a critical issue that strongly influences battery performance. In this work, first-principles density functional theory calculations are performed to investigate the interfacial properties of a solid-state battery system composed of Li metal anode and Li$_3$OCl solid electrolyte. The structural stability, electronic structure, and electrochemical behavior of the Li|Li$_3$OCl interface are systematically analyzed. Several interface orientations are constructed and compared in order to identify the most energetically favorable configuration. The electronic properties and interfacial charge redistribution are further examined to understand the nature of the interaction between Li metal and the Li$_3$OCl electrolyte. Our results indicate that the Li|Li$_3$OCl interface exhibits stable structural and electronic characteristics, with localized charge redistribution occurring near the interface region. The electrochemical stability against the insertion of an additional Li atom is also evaluated, showing that Li incorporation is energetically unfavorable in most layers of the electrolyte. These results suggest that the Li$_3$OCl electrolyte maintains good electrochemical stability in contact with Li metal. The present study provides atomic-scale insight into the interfacial behavior of Li|Li$_3$OCl and highlights the potential of Li$_3$OCl as a promising solid electrolyte for solid-state lithium batteries.
\end{abstract}




\maketitle

\section{Introduction}

The all-solid-state lithium battery using Li metal anode is a promising next-generation battery technology with higher energy density and significant safety advantages compared to conventional Li-ion batteries with liquid electrolyte \cite{takada2013progress,wang2015design,manthiram2017lithium}. Li-metal solid-state batteries have three main component: the anode, cathode, and electrolyte. The most common anode materials are lithium metal \cite{nagao2013situ}, lithium alloys \cite{sahu2014air} and graphite \cite{takada2003solid,takada2003compatibility}. The interfacial resistance at the electrolyte/anode and electrolyte/cathode interfaces is one of the paramount issues leading to poor performance of solid-state batteries. Interest in electrochemical cells based on solid-state electrolytes (SSE) dates back about 200 years \cite{bruce1997polymer}, and a rapid development of SSE emerged from 1960s. One of the most important problems that urgently to be overcome is how to enhance the ionic conductivity at the interfaces of the electrolyte/anode and electrolyte/electrode. Li|SSE interface has been studied by large-scale molecular dynamics which modeling Li diffusion mechanisms during Li stripping and plating and near the SSE interface \cite{yang2021interfacial}. Liu \textit{et. al.} \cite{liu2016interfacial} compared the interfacial energy and work of adhesion at different orientations which shows that Li(001) is the most energetically stable interfacial orientation and Li$_{2}$CO$_{3}$ shows better interfacial mechanical stability than LiF in solid electrolyte interphase (SEI).

The Li$^{+}$ interstitial is the most mobile Li$^{+}$ charge carrier in solid-state battery due to the lower diffusion barriers ($\sim$175 meV) \cite{stegmaier2017li+}. The dominance of a Li$^{+}$ charge carrier depends not only on its diffusion barrier but also on its concentration. What remains to be understood is how the Li|SSE interface affects the electrochemical energy of lithium batteries. It is generally acknowledged that the charger at the Li|SSE interface can remarkably influence the electrochemical performance of all-solid-state batteries (ASSBs). Currently, there is a lack of understanding of the role of Li|SSE interface and their mechanical and electrochemical properties.

A deeper understanding of the critical issues of SSB, especially interfacial challenges is requirement. However, there is a lack of understanding of the role of Li|SSE interfaces and their mechanical and electrochemical properties. Here, we present an interface between Li metal anode and Li$_{3}$OCl electrolyte to evaluate the electrochemical properties based on density functional theory (DFT). In this study, we perform first principles simulations to model the Li|SSE interface. Our work provides atomistic level understanding into the atomistic defects at the Li|SSE interface, and suggest future directions for atomistic-level interface engineering to improve the cycling stability of solid- state Li metal batteries. This knowledge is critical to understand the interface interactions/effects in Li batteries, and the electrochemical stability of Li|Li$_{3}$OCl interface with additional one Li-ion.


\section{Computational approaches}
All the calculations were performed with density functional theory (DFT) and projector augmented wave (PAW) \cite{blochl1994projector, kresse1999ultrasoft} pseudopotential, as implemented in the Vienna \textit{Ab-initio} Simulation Package (VASP) \cite{hafner2008ab} software. The generalized-radient approximation (GGA) of Perdew-Burke-Ernzerhof (PBE) \cite{perdew1996generalized,perdew1992atoms} is the exchange-correlation functional. A cut-off energy of 500 eV was used for the plane wave basis set, and the Methfessel-Paxton smearing approach was applied with the width of 0.2 eV. A supercell structure of 841 atoms was used for calculating the interfacial interactions and the corresponding energy value. An appropriate k-points sampling of Monkhorst-Pack grid with $1\times3\times3$ mesh was used to approximate the Brillouin zone. During the optimization of structure properties, atomic positions
were relaxed with the convergence criteria for energy and forces set to 10$^{-8}$ 
eV and 5$\times$10$^{-3}$ eV/\AA{} , respectively. Pseudo atomic calculations performed for Li, Cl and O are s$^{1}$p$^{0}$, s$^{2}$p$^{5}$, s$^{2}$p$^{4}$, respectively. The primitive cubic crystal structure of \ce{Li3OCl} antiperovskite is with pm3m symmetry: one oxygen atom occupies in the center of \ce{Li6} octahedra and Cl atoms occupy in the
eight corner of a cube. The calculated ground-state lattice parameter of \ce{Li3OCl} is 3.88 \AA{} which agreement with the experimental value 3.91 \AA{} at room temperature \cite{braga2014novel,zhao2012superionic}. The experimental lattice parameter of bcc \ce{Li2} is 3.51 \AA{} , and our first-principles calculated value is 3.44 \AA{} .
The interface structure of \ce{Li|Li3ClO} was built by BCC
\ce{Li2} metal including 175 atoms and \ce{Li3OCl} with 666 (Li = 396, O=126 and Cl = 144) atoms. Seven atomic layers of Li(100) slab and fifteen atomic layers of LOC(100) were used in the present work. In order to obtain the good lattice mismatch between the surface of bcc Li and \ce{Li3OCl}, the x- and y-axis of \ce{Li3OCl} are rotated 45\textdegree{}. The lattice mismatch is calculated as:

\begin{equation}
\delta=\Bigg|{\frac{S_{LOC}}{S_{Li}}-1}\Bigg|    
\end{equation}

where $S_{LOC}=(\sqrt{2} \times a_{LOC})^2$ and $S_{Li}$=a$_{Li}^2$
are the geometric unit cell areas of \ce{Li3OCl} and \ce{Li2} metal in the interface, respectively. The lattice mismatch $\delta$ between \ce{Li3OCl} and \ce{Li2} is 4.2$\%$, which greatly reduces the lattice incoherency strain and, hence, contributes to the stability of the interface. To avoid any spurious interfaces between the periodic replica, 16 \AA {} vacuums regions were introduced at the surfaces of BCC Li and \ce{Li3OCl}, respectively. We present a systematic study of the structural stability, electronic properties based on DFT, and properties a model with one extra Li atom. We also explore the effects of the interface on the performance of \ce{Li|Li3OCl} battery.

-------------------------------------------------------
-------------------------------------------------------
\section{Results and discussion}
\subsection{Structural and electronic properties of bulk Li and Li$_3$OCl}

\begin{figure}[ht!]
	\centering
		\includegraphics[scale=.50]{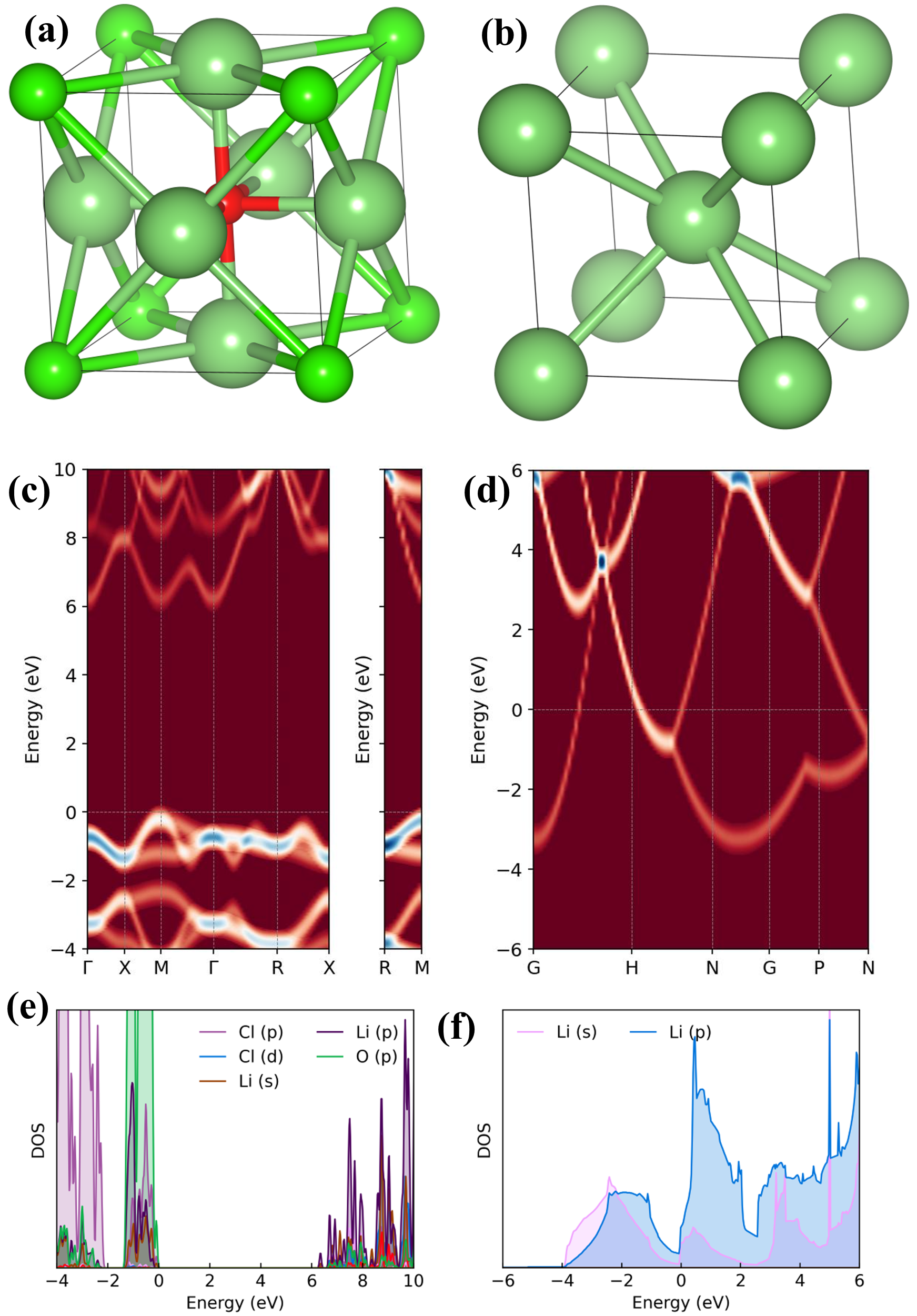}
	\caption{Fully optimized atomic structure of (a) anti-perovskite material Li$_3$OCl and (b) BCC Li system. The light green, dark green and red spheres are Li, Cl and O atoms, respectively. (c,d) k-resolved band structure calculations for the total density of states along the high-symmetry directions for Li$_3$OCl and Li, respectively. (e,f) Corresponding projected density of states of Li$_3$OCl and Li, respectively.}
	\label{FIG:1}
\end{figure}

Initially, we have fully optimized the structural and electronic properties of bulk Li BCC unit-cell and unit-cell of Li$_3$OCl antiperovskite material as shows in Figure \ref{FIG:1}(a-f). The optimized lattice parameters of Li$_3$OCl and Li are 3.88 \AA{}(a=b=c) and 3.44 \AA{} (a=b=c), respectively which is good consistent with previous work \cite{choi2022amorphisation}. The optimized bond lengths between Li-O and Li-Cl are 1.94 \AA{} and 2.95 \AA{}, respectively in Li$_3$OCl antiperovskite material and Li-Li bond length is 2.98 \AA{} in Li BCC system (see Figure \ref{FIG:1}a,b. Figure \ref{FIG:1}c,d displayed the electronic band structure of bulk Li$_3$OCl (LOC) and Li system. It was seen that the bulk \ce{Li3OCl} has an indirect band gap of 6.43 eV from valence band minimum (M-point) to conduction band maximum ($\Gamma$-point). The presence of a wide band gap indicates robust electrochemical stability in \ce{Li3OCl}, distinguishing it from numerous crystalline materials with comparable ionic conductivity \cite{wu2011effect}. From Figure \ref{FIG:1}d bulk Li system shows metallic character because some of the electronic states crosses the Fermi level.

The partial density of states (PDOS) of \ce{Li3OCl} (LOC) was calculated to examine the atomic contributions to its electronic structure, as shown in Figure \ref{FIG:1}e. The states near the Fermi level are mainly derived from O atoms, while the conduction band is dominated by Li states, indicating a predominantly ionic bonding character. To further validate this conclusion, Bader charge analysis \cite{henkelman2006fast} was performed. The calculated charges for Li, Cl, and O are $+0.94|e|$, $-1.83|e|$, and $-0.99|e|$, respectively, which are close to their nominal valence states. These results suggest that Li atoms are nearly fully ionized to Li$^+$, and the structure is stabilized by strong Coulombic interactions among Li$^+$, O$^{2-}$, and Cl$^-$ ions. Also, Figure \ref{FIG:1}f displayed the projected density of states of bulk Li which shows the metallic nature.

\subsection{Structural properties of Li|LOC interface}

We first study the interface configurations (Li|LOC) between Li metal anode and \ce{Li3OCl} solid-state electrolyte. In the LOC (100) slab, it can be considered as the combination of two different kinds of layers such as Li-O layer and Li-Cl layer. Also, based on the Li metal atomic arrangements, four potential interface structures are constructed between Li metal and LOC i.e. structure A, structure B, structure C and structure D, as shown in Figure \ref{FIG:2}. According to our calculations, the structure A is the stablest interface structure, while the structure B is equable with structure A which are shifted in structure A by 25\%. Li|LOC interface is disappeared in structure C and D. This is because the Coulombic attraction between O atom in Li-O layer and Li in Li metal anode at Li|LOC interface. In the work, the first-principles simulations are performed based on the structure A and the optimized interfacial distance are found to be 2.45 \AA{}. Which are good agreement with the previous work \cite{wu2020first,wu2020interfacial}.

\begin{figure}[ht!]
	\centering
		\includegraphics[scale=0.4]{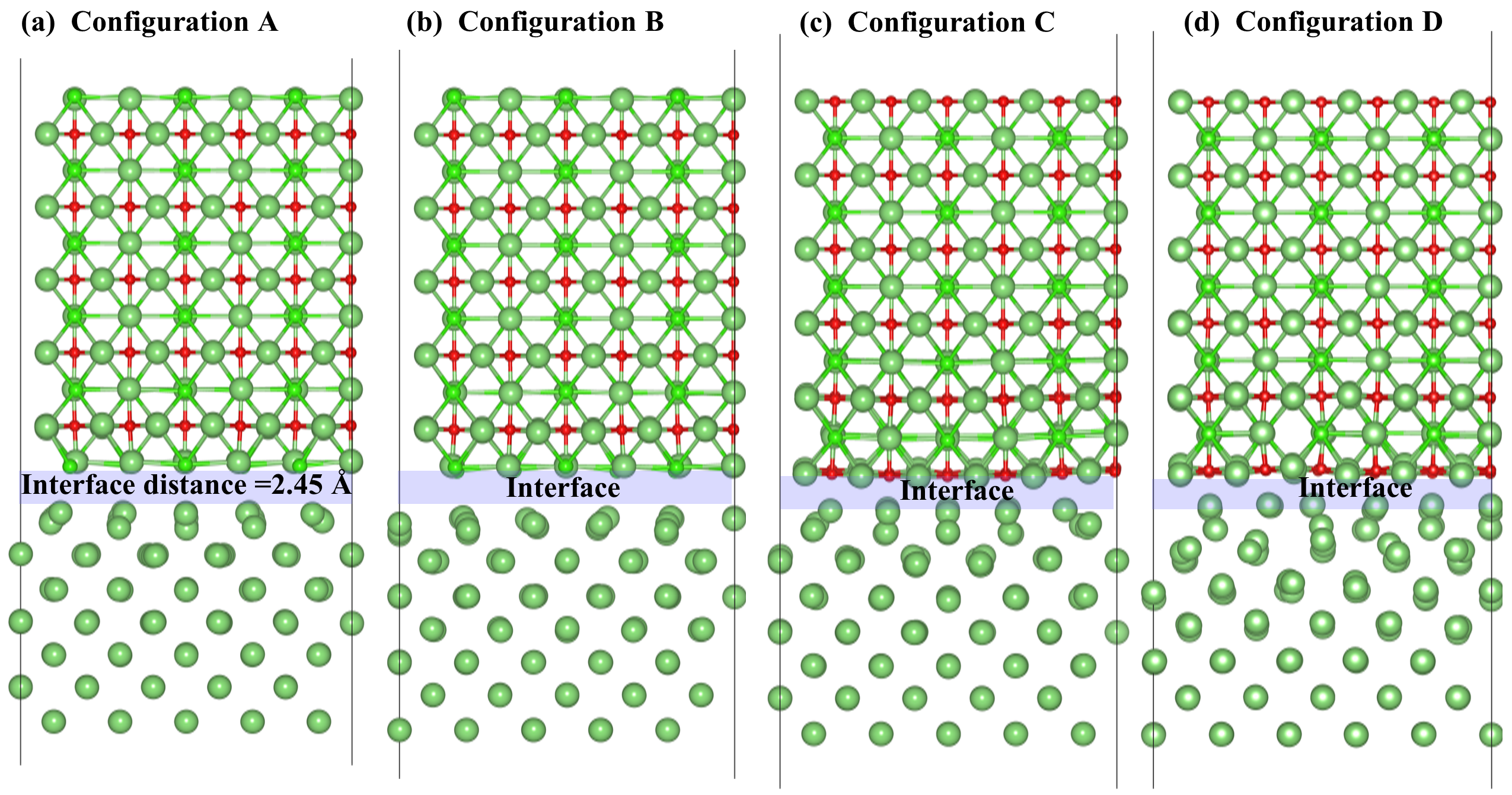}
	\caption{Fully optimized structural configurations of \ce{Li|LiOC} interface system. The light green, dark green and red spheres are Li, Cl and O atoms respectively.}
	\label{FIG:2}
\end{figure}

\begin{figure}
\centering
\includegraphics[scale=.58]{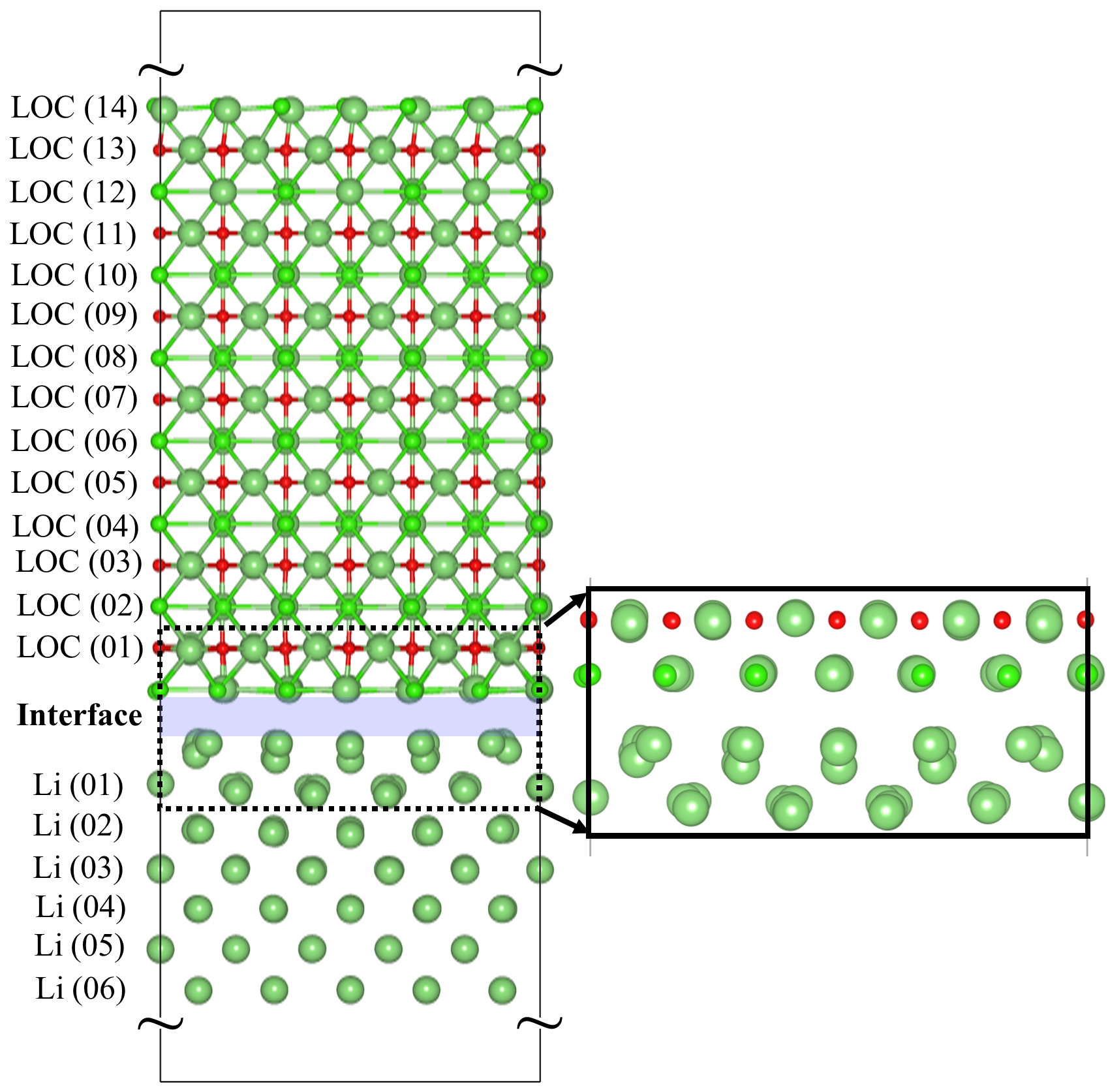}
\caption{The relaxed structure of configuration A of \ce{Li|Li3OCl} interface. The light violet color area is the interface. LOC(1), LOC(2), ... etc. present the first layer in the \ce{Li3OCl} side, the second layer in the \ce{Li3OCl} side, ....etc. Li(1), Li(2), ...etc present the first layer in the Li metal side, the second layer in the Li metal side, and so on. The inserted figure on the right side shows a zoomed-in atomic structure near to the interfacial region.}
\label{FIG:3}
\end{figure}

A clear understanding of the atomic structure at the interface is essential for explaining and improving the performance of solid-state batteries. During the relaxation of Li|LOC interface, inward or outward displacement of the atomic layers may occur under interface stress. Hence, gaining geometric structure information on the Li|LOC interface is one of the first steps necessary to determine interface properties. The fully optimized geometric configuration A is shown in Figure \ref{FIG:3}. There are small fluctuation in the relaxed geometric structure of \ce{Li3OCl} side at the interface Li-Cl atomic layer as shows in Figure \ref{FIG:3}. This is due to the high symmetry of the cubic \ce{Li3OCl} cubic structure and strong binding effect of the \ce{Li6O} octahedron. The BCC \ce{Li2} atomic positions around interface region
are changed significantly. The atoms are shifted about 0.60 \AA{} along the transverse +z-direction at the \ce{Li2} interface area. It as also seen that the Cl atoms are situated at same place i.e. there are no atomic displacement of Cl atoms in LOC sites. While Li atoms in LOC system are shifted 0.20 to 0.26 \AA{} along +z-direction due to the significant repulsion between Li-Li atoms from LOC and \ce{Li2} system.  The initial interficial distance was 3 \AA{} and the optimized interfacial distance is found to be 2.45 \AA{}. Our results are agreement with the previous reported work \cite{wu2020interfacial}. However, close to the interface region at BCC \ce{Li2} side, the variation of Li atomic positions depends on the atom types (Li or Cl) in the \ce{Li3OCl} interface region. This is because the Coulombic interactions between Li in \ce{Li2} and Li/Cl in \ce{Li3OCl} . Li atoms move large displacements is mainly due to the Colombic repulsion between Li ion on the \ce{Li2} anode and Li ion on the \ce{Li3OCl} electrolyte in the interface region. While other Li atoms slightly shifted due to the Coulombic attraction between Li atom in \ce{Li2} anode side and Cl atom in the \ce{Li3OCl} electrolyte side. As shown in Figure \ref{FIG:3}, different atomic layers are consisted as the Li|LOC interface. On the \ce{Li3OCl} side, Li-Cl layers and Li-O layers are alternately arranged. Li-Cl layer is closest to the interface.

\section{Electronic properties of Li|Li$_3$OCl interface}

\begin{figure*}[ht!]
	\centering
		\includegraphics[scale=1.1]{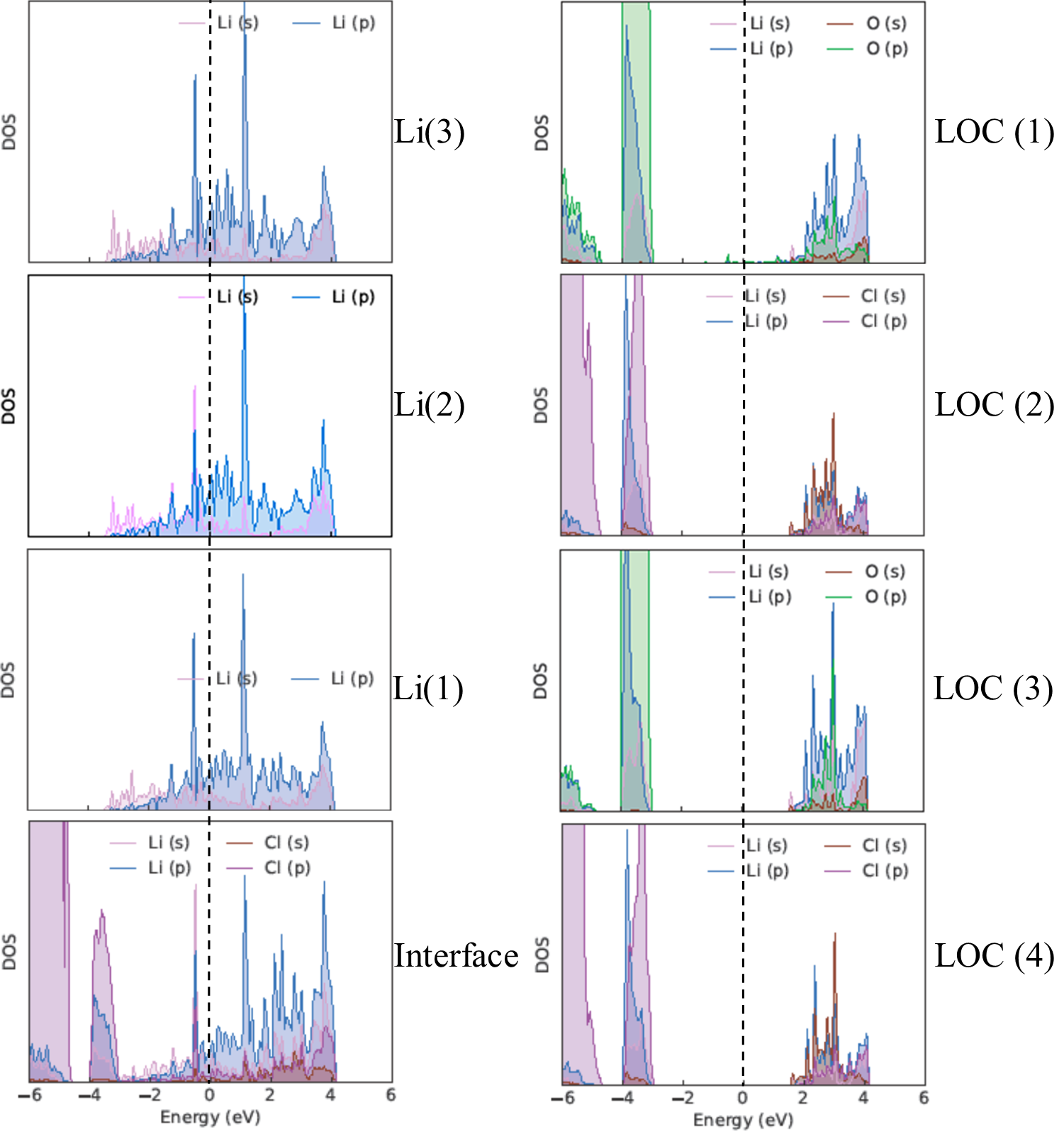}
	\caption{Partical and projected density of states of Li|LOC interface for Li battery at few layers of Li side as shown by Li (1), Li (2), Li (3) as well as few layers of LOC side LOC (1), LOC (2), LOC (3), LOC (4) including interfacial region which includes one LOC layer and one Li-layer.}
	\label{FIG:4}
\end{figure*}

To further understand the electronic behavior of the Li|Li$_3$OCl interface, we calculated the projected density of states (PDOS) for representative atomic layers across the interface, as shown in Fig.~\ref{FIG:4}. The PDOS analysis allows us to evaluate the electronic conductivity and possible electron transfer pathways at the metal–electrolyte junction. In the Li metal region, three representative layers are denoted as Li(1), Li(2), and Li(3), where Li(1) corresponds to the layer closest to the interface. On the electrolyte side, the “interface” refers to the first atomic layer Li-Cl of Li$_3$OCl directly adjacent to the Li metal slab.

The calculated PDOS indicates that, the Li metal slab retains its typical metallic character, with pronounced electronic states present at the Fermi level (E$_{F}$). In contrast, the Li$_3$OCl region exhibits insulating behavior. At the immediate interface on the Li$_3$OCl side, a small number of electronic states appear near the E$_{F}$, suggesting a localized metallic character at the first Li–Cl layer adjacent to the interface. However, this effect rapidly diminishes with increasing distance from the interface. The deeper layers of Li$_3$OCl display a clear band gap around the E$_{F}$, indicating that electrons cannot easily propagate into the bulk electrolyte. This behavior effectively suppresses electronic leakage from the Li metal anode into the electrolyte.

The total density of states for the Li|Li$_3$OCl system was calculated using the Perdew-Burke-Ernzerhof (PBE) generalized gradient approximation (GGA) functional. The resulting band gap of bulk Li$_3$OCl is approximately 4.94~eV, which agrees well with previous theoretical studies reporting a value of 4.84~eV \cite{kim2019predicting, stegmaier2017li+}. The hybrid functional calculations using HSE06 yield a larger band gap of 6.43~eV matches with previous literature \cite{kim2019predicting,choi2022amorphisation}. Despite this difference, both approaches confirm that Li$_3$OCl is a wide band gap insulator, which is beneficial for maintaining high electrochemical stability in solid-state battery applications. The slab models were fully relaxed with respect to both atomic coordinates and lattice parameters in order to obtain energetically stable interface structures. Structural relaxation reveals that the symmetry of the Li$_3$OCl lattice is partially modified near the interface due to the interaction with the Li metal slab. In particular, the atomic layers of Li$_3$OCl closest to the interface experience noticeable distortions compared to the bulk structure.

\begin{figure*}
\centering
\includegraphics[scale=0.92]{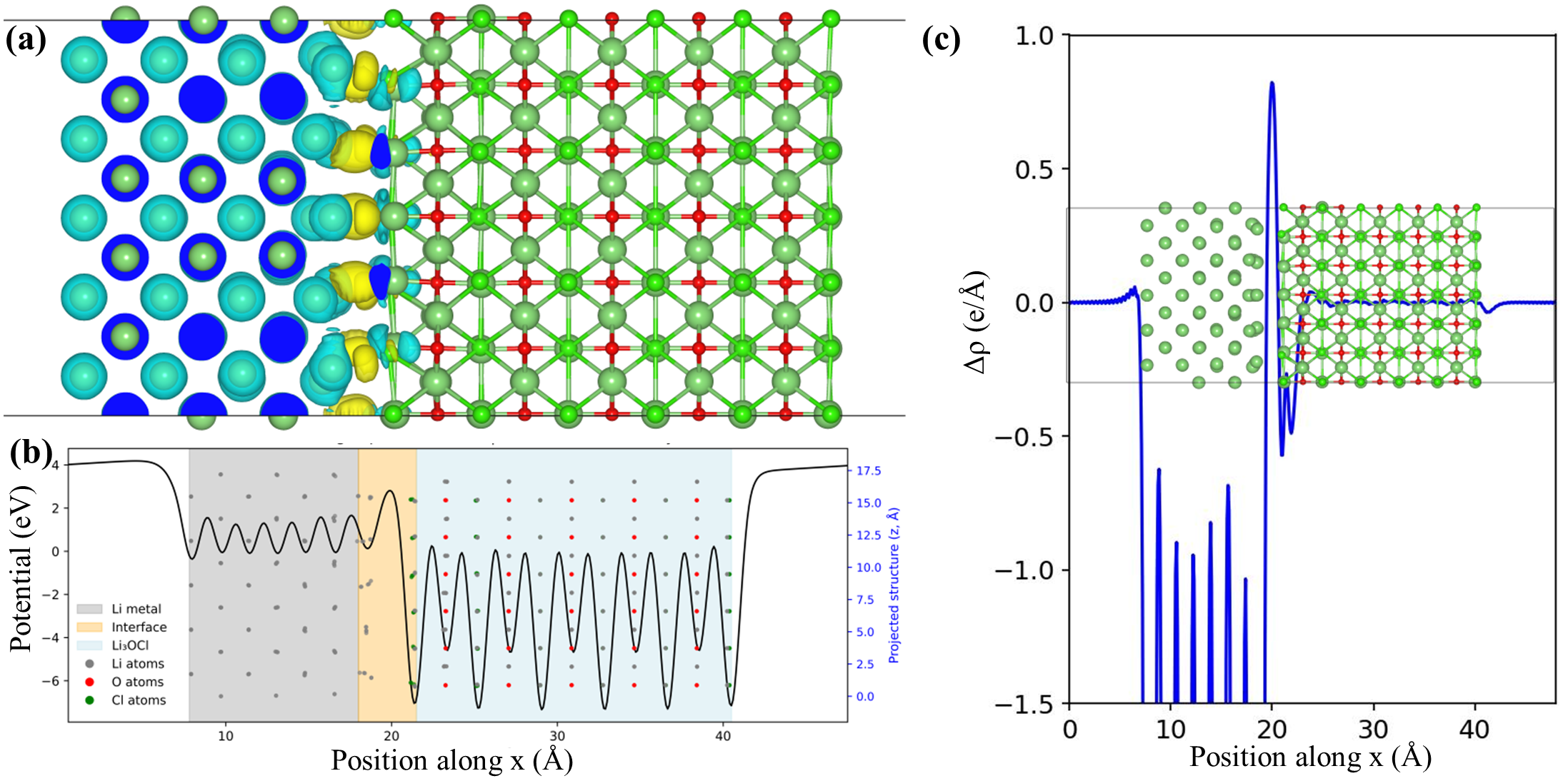}
\caption{Interfacial electronic structure of the Li|Li$_3$OCl system. (a) Charge density difference profile at the Li|Li$_3$OCl interface, where yellow and cyan isosurfaces represent electron accumulation and depletion, respectively, revealing localized charge redistribution near the interface. (b) Planar-averaged electrostatic potential profile along the interface normal direction ($x$-axis), showing the Li metal region, interfacial zone, and Li$_3$OCl electrolyte with a clear potential step arising from interfacial charge transfer. (c) Planar-averaged electron density difference $\Delta\rho(x)$ across the interface, indicating electron depletion on the Li metal side and accumulation within the first Li$_3$OCl layers. The charge redistribution is highly localized, confirming that the bulk Li$_3$OCl retains its insulating character while forming an interfacial dipole with Li metal.}
\label{FIG:5}
\end{figure*}

Figure~\ref{FIG:5} illustrates the interfacial electronic redistribution and electrostatic characteristics of the Li|Li$_3$OCl system. Figure \ref{FIG:5}(a-c) shows the charge density difference, the planar-averaged electrostatic potential profile along the interface normal, and the planar-averaged electron density difference $\Delta\rho(z)$, respectively. Figure~\ref{FIG:5}(a) shows the 3D charge density difference at the Li|Li$_3$OCl interface, which was obtained by subtracting the electron densities of the isolated Li metal slab and the Li$_3$OCl slab from that of the combined interface system (i.e. $\Delta \rho = \rho_{\ce{Li|Li_3OCl}}-\rho_{\ce{Li_3OCl}}-\rho_{Li}$). The yellow and cyan isosurfaces represent regions of electron accumulation and depletion, respectively. A pronounced redistribution of electronic charge is observed near the interface, indicating strong electronic interaction between the Li metal and the adjacent Li$_3$OCl layers. Electron depletion is primarily located near the Li atoms on the metal side, while electron accumulation is mainly distributed around the interfacial oxygen and chlorine atoms of the Li$_3$OCl lattice. This charge rearrangement suggests partial electron transfer from the Li metal toward the electronegative species in the electrolyte. However, the redistribution is largely confined to the first few atomic layers near the interface, indicating that the bulk of the Li$_3$OCl region remains electronically unaffected and preserves its insulating character.

The electrostatic potential profile along the interface normal direction is shown in Fig.~\ref{FIG:5}(b). The planar-averaged potential is plotted as a function of the position along the $x$-direction, which corresponds to the direction perpendicular to the interface. The shaded regions identify the Li metal slab, the interfacial region, and the Li$_3$OCl electrolyte. In the Li metal region, the potential remains relatively smooth due to the strong electronic screening characteristic of metallic systems. In contrast, the Li$_3$OCl region exhibits pronounced oscillations in the potential that correspond to the periodic atomic layers of the antiperovskite lattice. A clear potential step is observed across the interface, reflecting the formation of an interfacial dipole arising from charge transfer and structural relaxation. This potential variation indicates the presence of an internal electric field localized near the interface, which may influence lithium-ion transport and charge redistribution in the interfacial region.

Figure~\ref{FIG:5}(c) presents the planar-averaged electron density difference $\Delta\rho(x)$ along the same direction. Positive values of $\Delta\rho(x)$ correspond to electron accumulation, while negative values represent electron depletion. The profile reveals a sharp redistribution of charge in the vicinity of the interface, characterized by alternating peaks and valleys corresponding to accumulation and depletion layers. On the Li metal side, a strong negative peak indicates electron depletion near the surface layer of the metal. Immediately across the interface, a positive peak appears within the first Li$_3$OCl layer, confirming electron accumulation in the electrolyte region. Away from the interface, the amplitude of $\Delta\rho(x)$ rapidly decreases and approaches zero, demonstrating that the charge transfer is highly localized and does not extend deep into the electrolyte bulk. It means that, the combined analysis of charge density differences and electrostatic potential profiles reveals that the Li|Li$_3$OCl interface exhibits localized charge redistribution accompanied by the formation of an interfacial dipole. Importantly, the electronic perturbation is confined to the immediate interfacial layers, while the bulk Li$_3$OCl retains its insulating character. This behavior is favorable for solid-state battery applications because it suppresses long-range electron leakage from the Li metal electrode into the electrolyte while still allowing ionic transport across the interface.

\subsection{Insertion of one additional Li atom}

The performance of solid-state lithium batteries is strongly influenced by the concentration and mobility of charge carriers within the solid electrolyte. Among the possible charge carriers, lithium interstitials (Li$^{+}$) play a crucial role in determining ionic transport behavior. Therefore, understanding the interaction of Li interstitials with the Li|Li$_3$OCl interface is important for evaluating the electrochemical stability and lithium accommodation capability of the electrolyte near the metal electrode. To investigate this effect, an additional Li atom was introduced into different positions within the Li$_3$OCl region of the Li|Li$_3$OCl slab model. The total energies of the system were calculated after structural relaxation in order to evaluate the energetics associated with Li insertion at different atomic layers. Initially, the extra Li atom was placed at the interfacial layer on the Li$_3$OCl side. After relaxation, two typical behaviors were observed. In some configurations, the additional Li atom tends to migrate toward the Li metal region at the interface. In other cases, the inserted Li atom replaces a nearby Li atom in the interfacial layer, effectively pushing the original Li atom away from its lattice site. The relative stability of these configurations is determined by the electrochemical insertion energy associated with each Li position. The crystal structure of Li$_3$OCl consists of alternating atomic layers with different local chemical environments. In the Li$_3$OCl region, Li and Cl atoms occupy one set of layers, while Li and O atoms form the adjacent layers. Due to this layered arrangement, the available insertion sites for an additional Li atom depend strongly on the symmetry and coordination environment of the specific layer. In the even-numbered layers, two possible interstitial configurations exist because of symmetry considerations. One configuration corresponds to a higher symmetry insertion site, while the other arises from a slightly distorted local coordination environment. In contrast, the odd-numbered layers provide only a single type of insertion configuration because the oxygen atoms occupy distinct crystallographic positions.

The insertion of an additional Li atom was systematically examined up to the fifteenth layer on the Li$_3$OCl side of the interface. This depth was chosen because electronic and electrostatic interactions between the Li metal electrode and the electrolyte are expected to be strongest within the first several atomic layers near the interface, while deeper layers are progressively less affected by the presence of the metal. The local bonding environment around the inserted Li atom varies depending on the layer in which the interstitial is located. In even-numbered layers, the extra Li atom typically interacts with two oxygen atoms within the same layer. When placed at a high-symmetry interstitial site, the Li atom may coordinate with up to four surrounding oxygen atoms. In contrast, when the Li interstitial is introduced in an odd-numbered layer, it forms bonds with two oxygen atoms located in the neighboring upper and lower layers of the lattice.

In the pristine Li|Li$_3$OCl supercell, each oxygen atom forms an octahedral Li$_6$O coordination environment. When an additional Li atom is inserted, our calculations indicate the formation of a Li interstitial dumbbell configuration in some cases. In this configuration, the inserted Li atom shares a lattice position with an existing Li atom, effectively splitting the original Li site. Similar interstitial dumbbell configurations have previously been reported for bulk Li$_3$OCl \cite{emly2013phase}. Structural analysis further shows that when the additional Li atom is placed directly at the interfacial layer of Li$_3$OCl, the oxygen atoms maintain six Li-O bonds, which is identical to the coordination environment observed in the pristine structure. The corresponding Li–O–Li bond angles remain close to 90$^\circ$, indicating that the interface structure experiences only minor distortions upon Li insertion. However, when the extra Li atom is introduced into deeper layers of the electrolyte, the nearest oxygen atoms may form up to seven Li–O bonds. This increase in coordination slightly modifies the local geometry, leading to moderate distortions with Li–O–Li bond angles varying between approximately 86$^\circ$ and 94$^\circ$.

\section{Electrochemical energy}

The interfacial interaction between Li metal and the Li$_3$OCl electrolyte plays an important role in determining the electrochemical stability of the system. To evaluate the effect of excess lithium at the interface, we calculated the electrochemical insertion energy associated with introducing one additional Li atom at different atomic layers on the Li$_3$OCl side of the Li|Li$_3$OCl interface. In practical battery operation, lithium ions are generated at the Li metal anode during charging and migrate through the solid electrolyte toward the cathode. Therefore, the energetic stability of Li insertion within the electrolyte near the interface provides important insight into the interfacial electrochemical behavior of the system. The insertion energy was calculated from the total energies of the relaxed interface slab with and without an additional Li atom, together with the reference energy of bulk Li metal:

\begin{equation}
\Delta E = E_{Li|Li_{3}OCl+1Li} - \left(E_{Li|Li_{3}OCl} + \frac{1}{2}E_{Li_{2}}\right)
\end{equation}

where $E_{Li|Li_{3}OCl+1Li}$ is the total energy of the Li|Li$_3$OCl slab containing one extra Li atom, $E_{Li|Li_{3}OCl}$ represents the total energy of the pristine Li|Li$_3$OCl slab, and $E_{Li_{2}}$ corresponds to the energy of bulk Li metal. Within this definition, a positive value of $\Delta E$ indicates that Li insertion is energetically unfavorable and therefore electrochemically stable, whereas a negative value suggests that Li incorporation is energetically favorable and may lead to local reduction or structural instability.

\begin{figure*} 
\centering 
\includegraphics[scale=1.0]{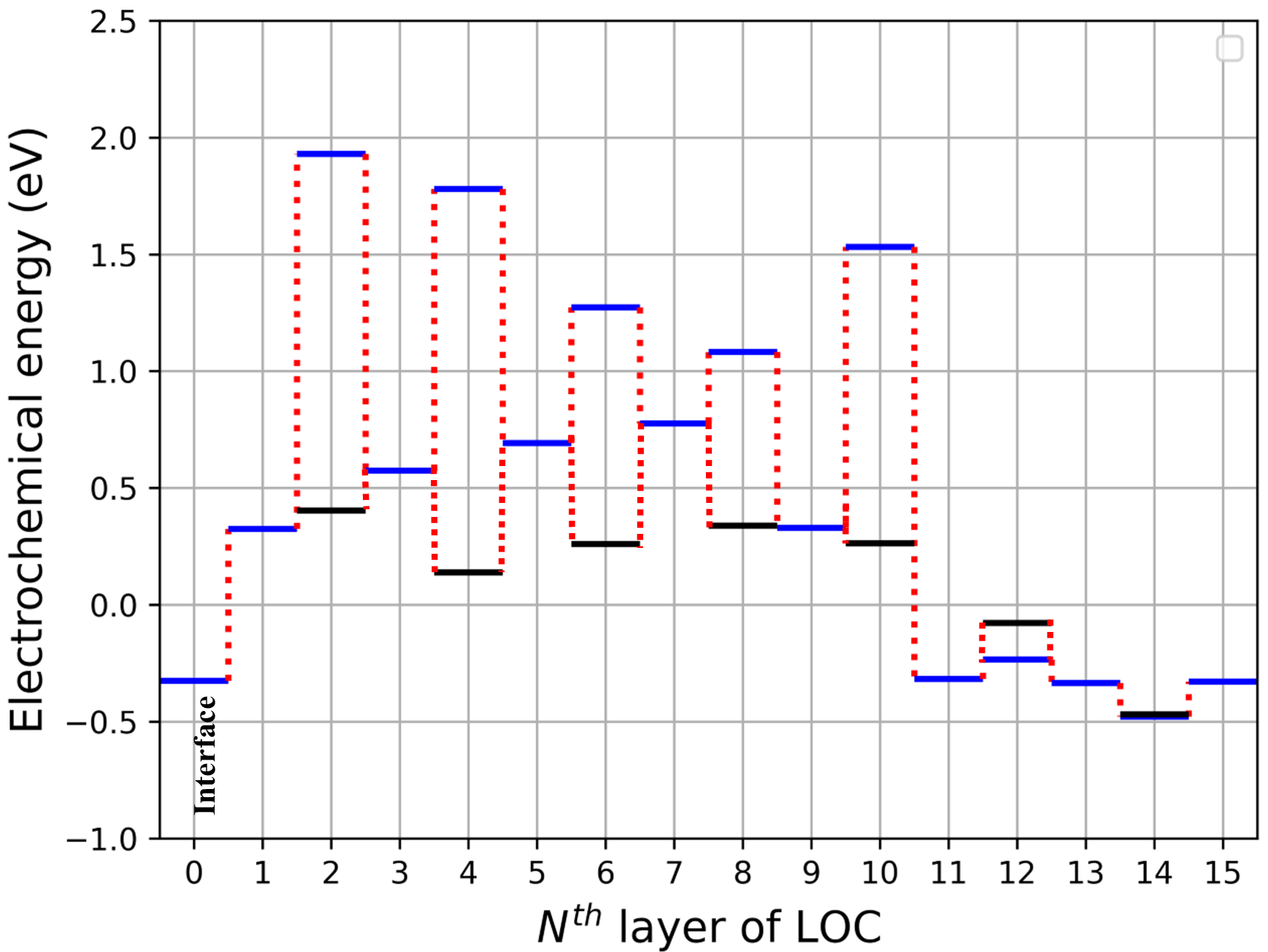} 
\caption{Electrochemical insertion energy of an additional Li atom at different atomic layers ($N$) on the Li$_3$OCl side of the Li|Li$_3$OCl interface. Positive energies indicate energetically unfavorable Li insertion and thus electrochemical stability, whereas negative energies suggest favorable Li incorporation near the interface. In some layers two energy values appear due to two possible Li insertion configurations arising from the local coordination environment.} 
\label{FIG:2} 
\end{figure*}

Figure~\ref{FIG:2} presents the calculated electrochemical insertion energies as a function of the atomic layer index on the Li$_3$OCl side of the interface. The layer index $N$ increases from the interface toward the bulk electrolyte region. Because of the layered structure of Li$_3$OCl, two distinct insertion configurations may exist in some layers depending on the local atomic coordination environment. In particular, layers containing both Li and Cl atoms provide two possible interstitial positions for the inserted Li atom, leading to two different insertion energies. In contrast, layers containing Li and O atoms allow only a single stable insertion configuration due to the unique coordination environment around oxygen atoms.

At the interface layer ($N = 0$), the insertion energy is slightly negative ($\Delta E \approx -0.33$ eV), indicating that Li insertion is energetically favorable at the immediate interface. This behavior suggests that the interfacial region is the most susceptible to lithium accumulation or structural rearrangement. The relatively low stability at the interface can be attributed to strong local interactions between Li atoms and the surrounding oxygen atoms, which stabilize the inserted Li atom and facilitate local charge transfer between Li metal and the electrolyte. Moving away from the interface, the electrochemical insertion energies become positive for most layers within the Li$_3$OCl region. For example, the second and fourth layers exhibit relatively large positive energies exceeding 1 eV, indicating that Li insertion at these positions is energetically unfavorable. These results suggest that the near-interface region of the Li$_3$OCl electrolyte maintains good electrochemical stability against lithium incorporation. In layers where two insertion configurations are possible, the calculated energies differ moderately, reflecting the influence of local symmetry and coordination environment on the stability of the inserted Li atom.

Interestingly, beyond approximately the tenth layer the calculated insertion energies become slightly negative again. However, the magnitude of these negative values remains relatively small (typically within a few tenths of an eV). This behavior can be attributed to intrinsic lithium interstitial defect formation within the Li$_3$OCl lattice rather than a direct destabilizing influence from the Li metal interface. In this deeper region of the electrolyte, the atomic environment approaches the bulk structure of Li$_3$OCl, where lithium interstitials can be weakly stabilized through local lattice relaxation or the formation of interstitial dumbbell configurations. Importantly, the negative energies observed in these deeper layers do not imply significant interfacial decomposition. Instead, they reflect the intrinsic defect chemistry of the Li$_3$OCl electrolyte. The influence of the Li metal electrode is therefore largely confined to the immediate interfacial region and does not propagate deeply into the electrolyte.

These results indicate that the Li|Li$_3$OCl interface exhibits a localized region of reduced electrochemical stability at the interface, while the majority of the Li$_3$OCl electrolyte maintains positive insertion energies and thus remains electrochemically stable against lithium penetration. This behavior is consistent with previous studies reporting negative insertion energies for Li interstitials in bulk Li$_3$OCl \cite{choi2022amorphisation}. Nevertheless, in the present interface model the destabilizing effect is limited to the first few atomic layers, suggesting that the bulk of the electrolyte preserves its structural and electrochemical integrity. These findings highlight the important role of interfacial atomic structure in determining lithium insertion energetics and provide valuable insight into the stability of Li$_3$OCl as a solid electrolyte in contact with Li metal.

\begin{figure} 
\centering 
\includegraphics[scale=0.46]{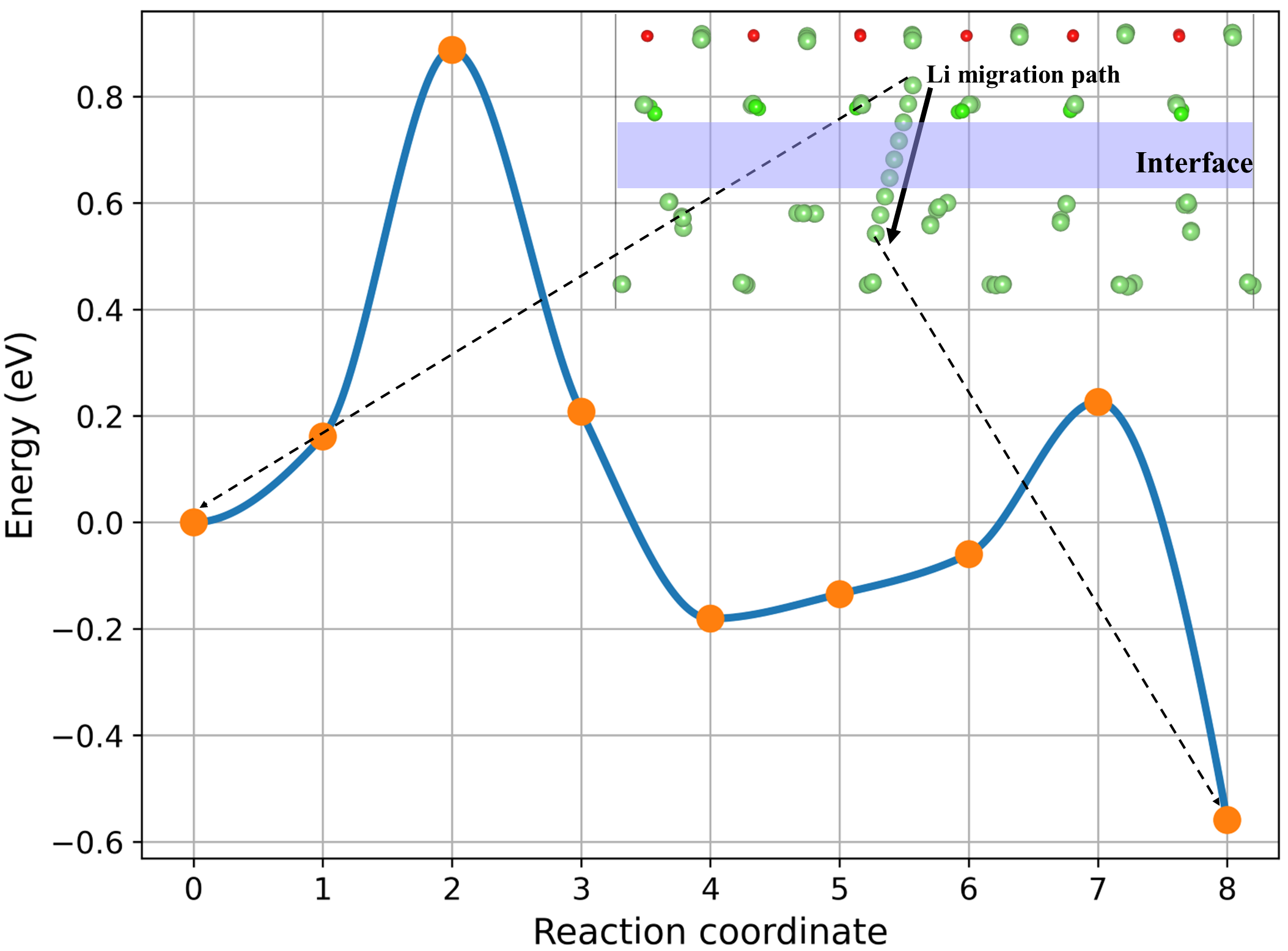} 
\caption{Minimum energy pathway for Li migration across the Li|Li$_3$OCl interface calculated using the nudged elastic band (NEB) method. The dots represent the discrete NEB images, while the solid curve is a spline interpolation to guide the eye. The calculated migration barrier is approximately 0.89 eV.} 
\label{FIG:7} 
\end{figure}

\subsection{Lithium migration across the Li|Li$_3$OCl interface}

To further understand lithium transport at the Li|Li$_3$OCl interface, the nudged elastic band (NEB) method was employed to determine the migration pathway of a Li atom from the Li$_3$OCl region toward the Li metal surface. The calculated minimum energy pathway reveals a migration barrier of 0.89 eV. The energy profile indicates that the final configuration, corresponding to Li located at the Li metal side, is energetically more stable than the initial state in the electrolyte by about 0.56 eV. This suggests that lithium atoms within the Li$_3$OCl region tend to migrate toward the metallic Li interface, reflecting the strong thermodynamic stability of Li in the metallic phase. The moderate migration barrier indicates that lithium transport across the interface is kinetically accessible, although not extremely fast \cite{wu2020interfacial,fu2023review}. Such behavior implies that Li$_3$OCl can support interfacial Li$^+$ transfer while maintaining overall electrolyte stability. These results provide further insight into the interfacial transport mechanism and highlight the role of the Li|Li$_3$OCl interface in regulating lithium plating and stripping processes in solid-state lithium batteries.

\section{Conclusions}
In this work, first-principles density functional theory calculations were performed to investigate the structural, electronic, and electrochemical properties of the Li|Li$_3$OCl interface for solid-state lithium batteries. The structural and electronic characteristics of bulk Li$_3$OCl and Li metal were first analyzed to establish the fundamental properties of the individual materials. Four possible interface orientations were constructed and compared, and the configuration with the lowest interfacial energy was identified as the most stable structure. Charge density difference, electrostatic potential, and planar-averaged charge density analyses reveal a noticeable redistribution of electronic charge near the interface, indicating electron transfer from the Li metal toward the Li$_3$OCl region. Electronic structure calculations further show that while the Li metal side retains its metallic nature, the Li$_3$OCl layers away from the interface preserve a wide band gap, confirming the insulating character of the electrolyte. The electrochemical stability of the interface was further examined by introducing an additional Li atom at different atomic layers within the Li$_3$OCl region. The calculated insertion energies demonstrate that Li incorporation is slightly favorable at the immediate interface but becomes energetically unfavorable for most layers inside the electrolyte. This behavior suggests that the destabilizing influence of Li metal is largely confined to the interfacial region, while the bulk Li$_3$OCl remains electrochemically stable. Our results also show that Li interstitials in the Li$_3$OCl electrolyte are stable when located near the Li|Li$_3$OCl interface, whereas Li-ion migration is unfavorable deeper within the solid electrolyte. These findings provide atomic-scale insight into interfacial interactions, lithium insertion behavior, and mechanical stability induced by interfacial species, highlighting the potential of Li$_3$OCl as a promising solid electrolyte for next-generation all-solid-state lithium batteries.


\subsection*{Declaration of competing interest}
The authors declare that they have no known competing financial interests or personal relationships that could have appeared to influence the work reported in this paper.

\section*{Acknowledgements}
D.S. and R.L. thank the Carl Tryggers Stiftelse for Vetenskaplig Forskning (CTS 22:2283) for financial support. R.L. would like also to thank the {\AA}forsk Foundation (grant number 22-206). The computations were enabled by resources provided by the National Academic Infrastructure for Supercomputing in Sweden (NAISS), partially funded by the Swedish Research Council through grant agreement no. 2022-06725. NAISS (2023/22-1367 and 2024/5-38) Sweden is acknowledged for providing computing facilities.


\bibliographystyle{unsrt}

\bibliography{Ref}

\end{document}